\documentclass[aps,prl,twocolumn,showpacs,superscriptaddress,letterpaper]{revtex4-1}
\usepackage{mathrsfs}
\usepackage{amsmath}
\usepackage{bm}
\usepackage{color}
\usepackage{lineno}
\usepackage{graphicx}
\usepackage{subfig}
\usepackage{hyperref}
\begin{document}
%\linenumbers
%\title{Breakdown of Adiabatic Invariance of Charged Particle Gyromotion under Low Frequency Perturbations}
%\title{Transition to Chaos in Charged Particle Gyromotion due to Low Frequency Perturbations}
%\title{Route to Chaotic Charged-Particle Gyromotion due to Low Frequency Perturbations}
\title{On Validity of Gyrokinetic Theory}
\author{Haotian Chen}\email[]{Email:chenhaotian@swip.ac.cn}
\affiliation{Southwestern Institute of Physics, Chengdu 610041, China}
%\affiliation{Center for Nonlinear Plasma Science and C.R. ENEA Frascati-C.P. 65, Frascati 00044, Italy}

\author{Liu Chen}
\affiliation{Institute for Fusion Theory and Simulation and School of Physics, Zhejiang University, Hangzhou 310027, China}
\affiliation{Department of Physics and Astronomy, University of California, Irvine, California 92697, USA}
\affiliation{Center for Nonlinear Plasma Science and C.R. ENEA Frascati-C.P. 65, Frascati 00044, Italy}
\author{Fulvio Zonca}
\affiliation{Center for Nonlinear Plasma Science and C.R. ENEA Frascati-C.P. 65, Frascati 00044, Italy}
\affiliation{Institute for Fusion Theory and Simulation and School of Physics, Zhejiang University, Hangzhou 310027, China}
\author{Jiquan Li}
\affiliation{Southwestern Institute of Physics, Chengdu 610041, China}
\author{Min Xu}
\affiliation{Southwestern Institute of Physics, Chengdu 610041, China}

%\author{Haotian Chen}
%%\email[Correspondence and requests for materials should be addressed to H. C., email: ]{chenhaotian@swip.ac.cn}
%\email[email: ]{chenhaotian@swip.ac.cn}
%%\email[Correspondence and requests for materials should be addressed to H. C., email: ]{haotianchen-ext@us.es}
%\affiliation{Southwestern Institute of Physics, Chengdu 610041, China}

\vskip 0.25cm

\date{\today}

\begin{abstract}

	We study the validity of gyrokinetic theory by examining the destruction of magnetic moment adiabatic invariant in the presence of fluctuations.
	Contrary to common assertions, it is shown for the first time that  the gyrokinetic theory rests not only on the magnetic moment  conservation, but also on the fact that the particle dynamics constitutes a boundary layer problem.
	For low frequency fluctuations, there exists a quantitative, frequency independent threshold below which the adiabaticity is preserved, allowing thereby the general validity of gyrokinetic theory.
	The adiabaticity threshold in the high frequency regime, however, depends sensitively on frequency, which questions the generalization of gyrokinetic equation to arbitrary frequencies.
	Further analyses suggest that it is not feasible  to construct a reduced kinetic equation based on superadiabaticity.
	%These findings offer a physics basis for gyrokinetic theory.
	%These findings offer physical insights into the  basis of gyrokinetic theory.
%These findings provide useful insights into the physics basis of gyrokinetic theory.

\end{abstract}


\maketitle
%\begin{linenumbers}

%\begin{linenumbers}
%\end{linenumbers}

%\section{Introduction}
%\label{sec:may:01:16:01}
%\emph{Introduction.}---
Gyrokinetics, as one of the major achievements of modern plasma physics, has a profound impact on the understanding of magnetized plasmas in laboratory devices and nature  \cite{brizard07,krommes12}.
Historically, it was established on the basis of an asymptotic construction of the magnetic moment adiabatic invariant $\bar{\mu}$, initially through the gyro-averaging method \cite{taylor67, rutherford68, taylor68, catto, frieman} and later through  the Lie-transform perturbation theory \cite{littlejohn79,dubin83,hahm88, brizard89, sugama}.
Hence gyrokinetics is essentially a reduced kinetic theory derived from adiabaticity,  in which the details of the charged particle's gyromotion are not of dynamical importance.
%Along with advances in the high-performance computing and numerical algorithms, 
Along with advances in high-performance computing and analytical techniques, 
%gyrokinetics has made a deep and long-lasting impact on  plasma physics research. 
gyrokinetics is now widely recognized as the standard theory  of low frequency (compared to the gyrofrequency) plasma phenomena. It serves as the basis for numerous simulation codes and theoretical models used to study plasma instabilities, turbulence, and transport processes \cite{brizard07, krommes12, horton, kikuchi, chen16}.
%As an illustrative example in this regard, an ambitious simulation project
%has been launched recently to deliver a high-fidelity whole device model of tokamak fusion devices within the gyrokinetic framework \cite{ab}.
	An illustrative example in this regard is the recently launched ambitious simulation projects aimed at delivering a high-fidelity whole device model of tokamak plasmas within the gyrokinetic framework (see, e.g., Ref. \cite{ab}).

%However, despite the practical success of gyrokinetic theory, its validity, or equivalently the existence of adiabatic invariant,   is typically assessed by a qualitative estimate  of  the so-called gyrokinetic ordering \cite{frieman, brizard07} and often assumed. 
%However, despite the practical success of gyrokinetic theory, its validity is conventionally considered synonymous with the existence of adiabaticity, and typically assessed by a qualitative estimate  of  the so-called gyrokinetic ordering \cite{frieman, brizard07}. 
%In particular, previous studies mainly focus on the construction of adiabatic invariant \cite{kruskal,taylor67,dubin82, qin06}, there is relatively little demonstrated understanding of its validity limits in the literature.
However, despite the practical success of gyrokinetic theory, its validity is typically assessed by a qualitative estimate  of  the so-called gyrokinetic ordering \cite{frieman, brizard07} and often assumed. 
In particular, while it is frequently quoted that the validity of gyrokinetic theory is synonymous with the existence of adiabaticity, previous studies mainly focus on the construction of adiabatic invariant \cite{kruskal,taylor67,dubin82, qin06}, there is relatively little demonstrated understanding of its validity limits in the literature.
%In particular, while the validity of gyrokinetic theory is conventionally considered synonymous with the existence of adiabaticity, previous studies mainly focus on the construction of adiabatic invariant \cite{kruskal,taylor67,dubin82, qin06}, a thorough discussion of its destruction is still lacking.
The aim of this work is to illuminate what kind of physics sets the validity  limits of gyrokinetic theory.
%The aim of this Letter is to illustrate how the adiabatic invariant $\bar{\mu}$ would be broken and accordingly the gyrokinetic theory would become invalid.
Below, we will investigate (i) under which conditions the adiabaticity would be broken and accordingly the gyrokinetic theory would become invalid; (ii) why the high frequency gyrokinetic theory, which was developed in the early 1980s \cite{chen83b,tsai84, qin99, qin00, kolesnikov,yu}, has not gained much  popularity in implementation; 
(iii) whether the superadiabaticity \cite{rosenbluth72} could also be utilized to construct a reduced kinetic theory.
%(iii) whether the superadiabaticity, which was discovered even before the gyrokinetic theory \cite{rosenbluth72},  could also be utilized to construct a reduced kinetic theory.

%\section{Theoretical Model}
%\label{sec:model}
%%\subsection*{Gyrokinetic theory of zonal flow excitation}
%\subsection*{Gyrokinetic Model}
%\label{sec:may:01:16:02}
%\emph{Gyrokinetic model.}---
%For simplicity and hence clarity, 
%As a first step in addressing the validity of gyrokinetic approach in long timescale simulations, we investigate the $\bar{\mu}$-conservation in the presence of electric and magnetic perturbations in the simplest nontrival case. 
%For clarity, we restrict ourselves in the present study to the simplest nontrivial case.
%For clarity of the physics presentation, we restrict our attention to the simplest nontrivial case.
%we adopt the Taylor's model \cite{taylor67} as our prototype.  

To elucidate the key physics involved in these highly complex problems, we restrict our attention to the Taylor's model \cite{taylor67}.
%To elucidate the key physics involved in these highly complex problems, we restrict our attention to the Taylor's model \cite{taylor67} that provided a seminal example for the modern development of gyrokinetic theory \cite{brizard07}.
%To elucidate the key physics involved in these highly complex problems, we restrict our attention to the Taylor's model \cite{taylor67} that facilitated the modern development of gyrokinetic theory \cite{brizard07}.
%To elucidate the key physics involved in these highly complex problems, we restrict our attention to the Taylor's model \cite{taylor67} that motivates the modern development of gyrokinetic theory \cite{brizard07}.
%For clarity of the physics presentation, we restrict our attention to the Taylor's model \cite{taylor67}.
 This paradigm concerns the charged particle motion in a uniform magnetic field $\bm{B}=B\bm{e}_{z}$ with transverse perturbations, which, after a straightforward derivation, is described by the Hamiltonian 
\begin{eqnarray}
\label{eq:motion0}
	H=\frac{1}{2}(p^{2}+q^{2})+A\cos\omega t\cos q.
	%\dot{q}&=&p\nonumber\\
	%\dot{p}&=&-q+A\cos\omega t \sin q.
\end{eqnarray}
Here 
$\omega$ is the wave frequency and time is normalized to the gyroperiod $\Omega^{-1}=mc/eB$, with $e$ being the charge and $m$ the mass.
Though simple, Taylor's model is of  fundamental importance  in magnetized plasmas and was a seminal example for the modern development of gyrokinetic theory \cite{brizard07}.
It represents the simplest paradigm for wave-particle dynamics in 
both the electrostatic drift wave (with $A=-k_{\perp}^{2}e\delta\phi/(m\Omega^{2})$) \cite{taylor67} and shear Alfv\'{e}n wave (with $A=(k_{\perp}v_{A}/\Omega)(\delta B_{\perp}/B)$) \cite{chen01},
where $k_{\perp}$ is the perpendicular wavenumber,  $v_{A}$ denotes the Alfv\'{e}n velocity, and $\delta \phi$ and $\delta B_{\perp}$ are, respectively, amplitudes of the fluctuating electrostatic potential and magnetic field.
Note also that the nonlinearity parameter $A$ is determined by the spatial scale and amplitude of the perturbation, but is independent of the wave frequency.
%From Eq. (\ref{eq:motion0}) we note that the charged particle motion can be composed of two parts, a gyromotion due to the background magnetic field and a wave-particle trapping introduced by the perturbation.
%Therefore, the motion in $q$ is equivalent to a one-dimensional simple harmonic oscillator (SHO) driven by an external field with the amplitude $A=-k^{2}e\phi/(m\Omega^{2})$, and
The particle dynamics in Eq. (\ref{eq:motion0}) is composed of two parts: the gyromotion about the background magnetic field (the $q^{2}/2$ term),  and the wave-particle trapping due to the perturbation (the $A\cos\omega t\cos q$ term).
 One can easily verify that the phase space flow arising from Eq. (\ref{eq:motion0})  exhibits time reversal symmetry.
Whenever $(q(t), p(t))$ is an orbit of the system with the initial condition $(q_{0}, p_{0})$, both $(-q(-t), p(-t))$ and $(q(-t), -p(-t))$ will be physically allowable orbits, with the initial conditions $(-q_{0}, p_{0})$ and $(q_{0}, -p_{0})$, respectively. 
%two symmetries. One is the reflection symmetry that  the orbits lie symmetrically about the $p$-axis, i.e., if $(q(t), p(t))$ is a  solution of Eq. (\ref{eq:motion0}),  then  $(-q(-t), p(-t))$ is still a solution. 
%Theoretically, Eq. (\ref{eq:motion0}) can be seen as a dynamical system formally obtained  from the Hamiltonian
%\begin{eqnarray}
%\label{eq:hamiltonian1}
	%\tilde{H}=\frac{1}{2}(p^{2}+q^{2})+A\cos\omega t\cos q.
%\end{eqnarray}
The required invariant for periodic orbits  can be formulated in terms of the action integral as $I=\oint_{\gamma}pdq$,
%\begin{eqnarray}
%\label{eq:action0}
	%I=\oint_{\gamma}pdq,
%\end{eqnarray}
%where $\gamma$ is a closed path obtained by following the flow in phase space. 
where $\gamma$ is a complete period of the orbit at a constant time.
%With an infinitesimal perturbation ($A\to 0^{+}$), one readily retrieves the magnetic moment invariant given in \cite{taylor67} 
%\begin{eqnarray}
%\label{eq:action1}
	%%\frac{I}{2\pi}=\frac{1}{2}(p^{2}+q^{2})+A[\cos q-J_{0}(\sqrt{p^{2}+q^{2}})],
	%%\frac{I}{2\pi}=\frac{1}{2}(p^{2}+q^{2})+A\cos\omega t[\cos q-J_{0}(\sqrt{p^{2}+q^{2}})],
	%I=\pi(p^{2}+q^{2})+2\pi A\cos\omega t[\cos q-J_{0}(\sqrt{p^{2}+q^{2}})],
%\end{eqnarray}
%with $J_{0}$ being the Bessel function.

Further progress is possible if one introduces a variable $h$ conjugate to $t$, yielding the extended phase space Hamiltonian
\begin{eqnarray}
\label{eq:hamiltonian2}
	\mathcal{H}=\frac{1}{2}(p^{2}+q^{2})+A\cos \omega t\cos q-h.
\end{eqnarray}
In this way, the time-dependent one-dimensional Hamiltonian system is replaced by a two-dimensional Hamiltonian system where time plays a role analogous to that of an angle variable. 
%Before proceeding with the specific analysis of Eq. (\ref{eq:hamiltonian2}), it is illuminating to note that 
%Technically, to illustrate the long time-scale ($\omega t\gg  1$) behavior of the Taylor's model, we can construct the Poincar\'{e} map of the system (\ref{eq:hamiltonian2}) 
Technically, to illustrate the long time-scale ($\omega t\gg  1$) dynamical complexity of the system, we construct the Poincar\'{e} map of the flow in extended phase space:
	$\bm{x}_{n+1}=\mathcal{M}_{T}\bm{x}_{n}$,
%Therefore, since we are mainly interested in the dynamics on a long time scale $\omega t\gg  1$, an area-preserving mapping of surface of section is constructed  as
%\begin{eqnarray}
%\label{eq:map}
	%\bm{x}_{n+1}=\mathcal{M}_{T}\bm{x}_{n},
%\end{eqnarray}
where $\bm{x}=(q, p)$, and the mapping $\mathcal{M}_{T}$  is defined  such that the point $\bm{x}_{n}$ at $t_{n}$ is advanced by Eq. (\ref{eq:hamiltonian2}) to the next crossing point $\bm{x}_{n+1}$ at $t_{n+1}=t_{n}+T$, with $T=2\pi/\omega$. 

\begin{figure}[!htp]
\centering
\includegraphics[scale=0.4]{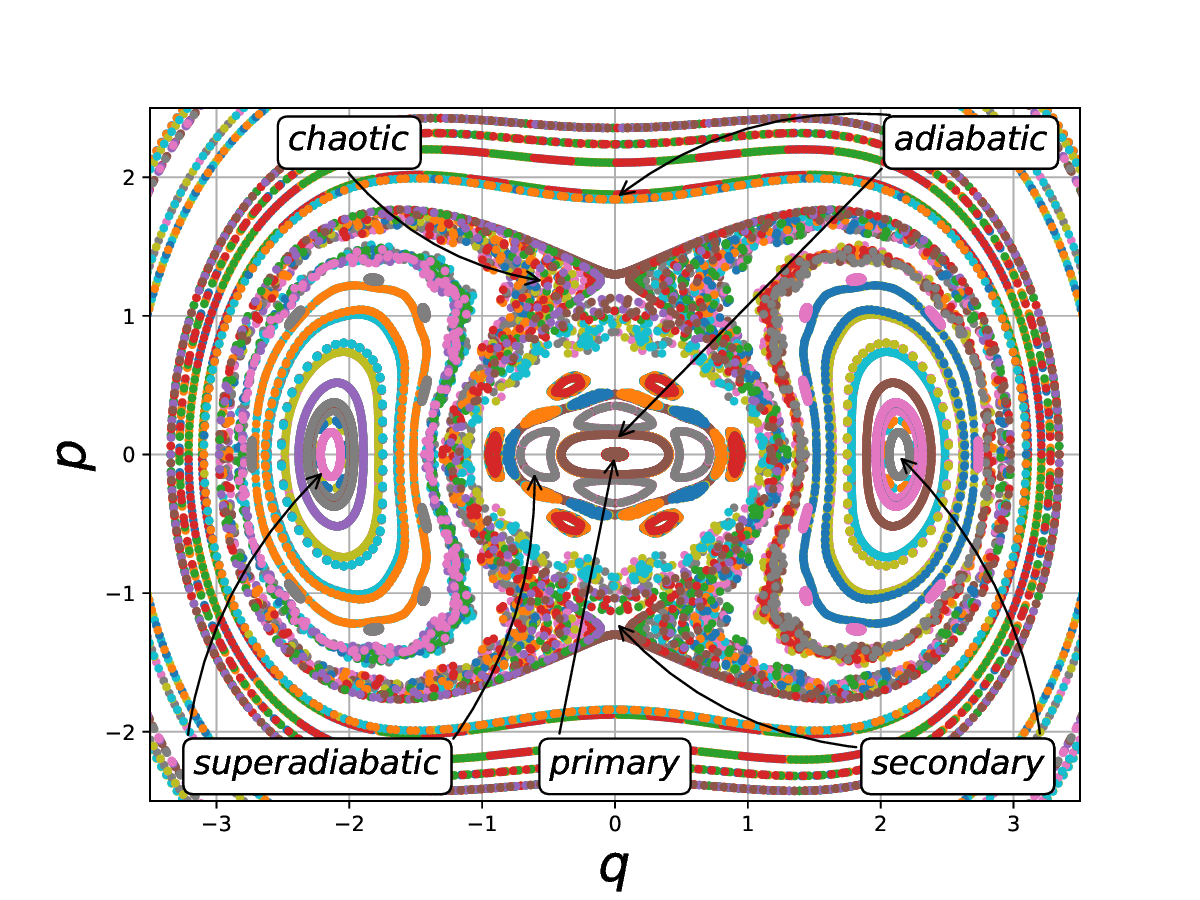}
	\caption{Phase space portrait of $\mathcal{M}_{T}$ for $\omega=0.5$ and $A=1.3$. Examples of  (primary and secondary) fixed points  and  (adiabatic, superadiabatic and chaotic) orbits  are shown explicitly.}
\label{eps:orbits}
\end{figure}

%\section{Stability Analyses}
%\label{sec:stability}
In Hamiltonian systems the transition to chaos can be understood looking at the behavior of orbits close to fixed points, which are either  elliptic (stable) or  hyperbolic (unstable).
As depicted in Fig. \ref{eps:orbits}, $\mathcal{M}_{T}$ has two different types of fixed points in accordance with the significance of nonlinearity. The primary fixed point $\bm{x}_{p}=(0,0)$ is also the fixed point of the original system (\ref{eq:motion0}). It becomes unstable when the linear parametric resonance \cite{arnold} occurs. Secondary fixed points $\bm{x}_{s}$, as assured by Poincar\'{e}-Birkhoff theorem \cite{lichtenberg},   are nonlinearly generated by a finite amplitude perturbation, half of which are elliptic and the other half hyperbolic in an alternating sequence. 
%In general, the stability of secondary fixed points is not amenable to analytical analysis and numerical computations are mandatory.
In general, the stability of secondary fixed points is not amenable to analytical analysis, numerical computations are mandatory.
%\end{linenumbers}

%\begin{figure}[!htp]
%\centering
%\includegraphics[scale=0.50]{fixedpoints1.eps}
	%\caption{(Color online) Phase space portrait of $\mathcal{M}_{T}$ for $\omega=0.5$ and $A=0.8$. Examples of fixed points are indicated explicitly.}
%\label{eps:fixedpoints}
%\end{figure}
\begin{figure}[!htp]
\centering
\includegraphics[scale=0.45]{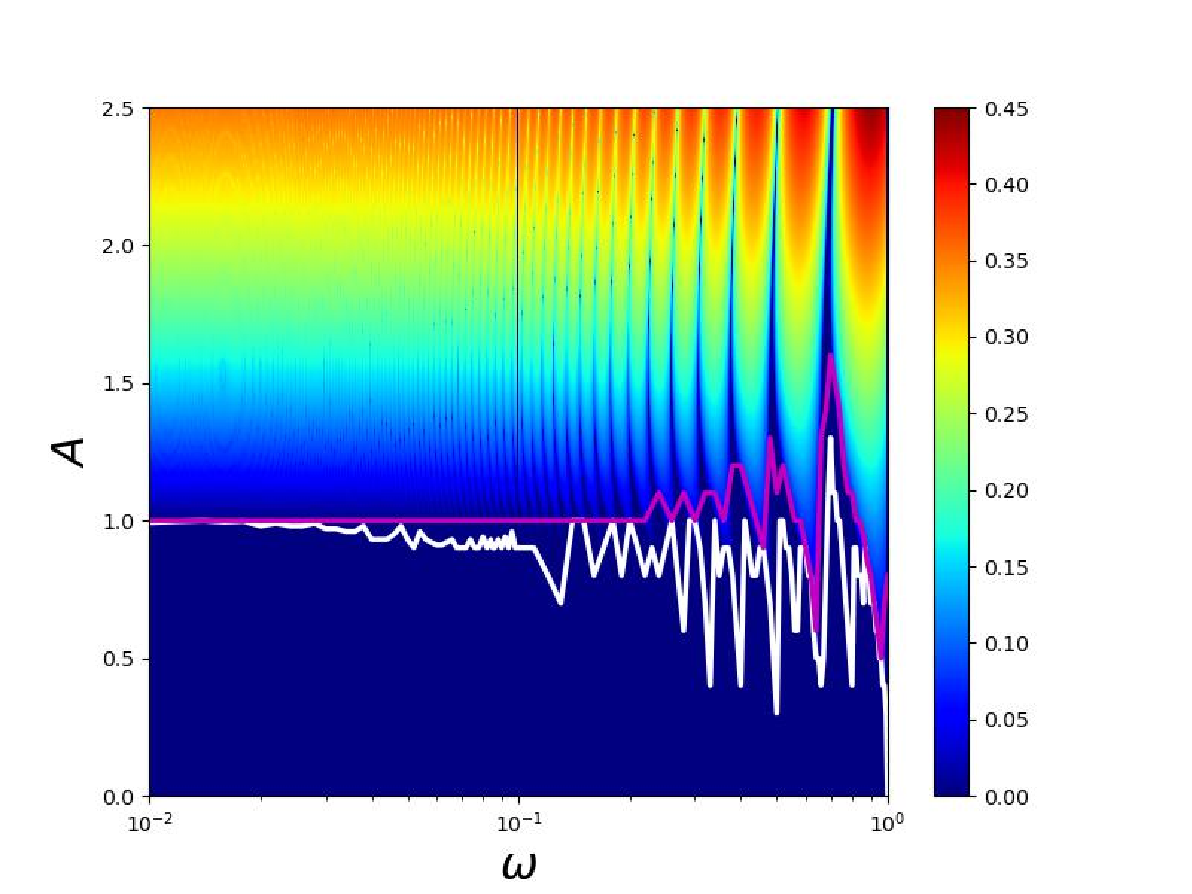}
	\caption{$\textrm{Im}\lambda$ contours in the $\omega-A$ plane. The white (magenta) line represents the threshold numerically evaluated  for the onset of superadiabatic (chaotic) orbits.}
\label{eps:lambda}
\end{figure}

%\begin{linenumbers}
The nature of the primary fixed point, however, can be studied analytically  by taking $\cos q\simeq 1-q^{2}/2$ in  Eq. (\ref{eq:motion0}).
The equation of motion reduces to a Mathieu equation:
%which yields a Mathieu equation \cite{chen01}:
%Further analytic progress in elucidating the nature of the principal fixed point is possible if we take the limit $\sin q\simeq q$ in Eq. (\ref{eq:motion0}),  
\begin{eqnarray}
\label{eq:mathieu}
	\ddot{q}+(1-A\cos\omega t)q=0.
\end{eqnarray}
According to the Floquet theory, the formal solution of Eq. (\ref{eq:mathieu}) can be expressed as
%in terms of the Fourier decomposition
%$q=\textrm{exp}(i\lambda t)\sum_{n}Q_{n}\textrm{exp}(i n \omega t)$,
\begin{eqnarray}
\label{eq:mathieu_solution}
	%q=e^{\frac{i}{2}\lambda\omega t}\sum_{n}Q_{n}e^{i n \omega t},
	q=e^{i\lambda t}\sum_{n}Q_{n}e^{i n \omega t},
\end{eqnarray}
where the Floquet characteristic exponent $\lambda$ is a measure of the rate of separation of orbits close to the primary fixed point. 
$\bm{x}_{p}$ is elliptic (hyperbolic) if $\lambda$ is real (complex).
Substituting Eq. (\ref{eq:mathieu_solution}) into Eq. (\ref{eq:mathieu}) and following the procedure in Ref. \cite{mclachlan}, one obtains
\begin{equation}
\label{eq:lambda}
	\lambda=\begin{cases}
		\frac{\omega}{\pi}\arcsin\sqrt{D(0)\sin^{2}(\frac{\pi}{\omega})},	& \quad \omega\ne \frac{1}{l},\\
		\frac{\omega}{2\pi}\arccos[2D(\frac{\omega}{2})-1], & \quad \omega=\frac{1}{l},
	\end{cases}
\end{equation}
where  $D$ is the determinant of an infinite tridiagonal matrix $\mathbf{D}=(d_{jk})$ with
%\begin{widetext}
%\begin{displaymath}
	%D(\lambda)=
	%\textrm{det}\left( \begin{array}{ccccccc}
		%\ldots & \ldots &\ldots & \ldots &\ldots &\ldots&\ldots\\
			%\vdots &  1& \frac{A}{2[(\lambda-2\omega)^{2}-1]} & 0 & 0 & 0 &\vdots \\
			%\vdots & \frac{A}{2[(\lambda-\omega)^{2}-1]} &1 & \frac{A}{2[(\lambda-\omega)^{2}-1]} & 0 & 0 &\vdots \\
			%\vdots & 0 & \frac{A}{2[\lambda^{2}-1]} &1 & \frac{A}{2[\lambda^{2}-1]} &  0 &\vdots \\
			%\vdots & 0& 0 & \frac{A}{2[(\lambda+\omega)^{2}-1]} &1 & \frac{A}{2[(\lambda+\omega)^{2}-1]} &  \vdots \\
			%\vdots & 0&  0&0 & \frac{A}{2[(\lambda+2\omega)^{2}-1]} &1 &  \vdots \\
			%\ldots & \ldots&  \ldots&\ldots & \ldots &\ldots &  \ldots 
	%\end{array}\right).
%\end{widetext}
\begin{eqnarray}
\label{eq:D}
	d_{jk}=\begin{cases}
		1 & \textrm{if $k=j$,}\\
		\frac{A}{2[(\lambda+j\omega)^{2}-1]} & \textrm{if $k=j\pm1$,}\\
		0 & \textrm{otherwise.}
	\end{cases}
\end{eqnarray}
$l, j, k$ are integers.
For sufficiently small $A\ll \omega^{2}/2$, an analytic expression for $D(0)$ can be concisely given as $D(0)\simeq 1-RA^{2}$ with 
	$R\equiv (\pi/\omega)\cot(\pi/\omega)/2(4-\omega^{2})$ \cite{mclachlan}.
%\begin{eqnarray}
%\label{eq:r}
	%R\equiv \frac{1}{2(4-\omega^{2})}\frac{\pi}{\omega}\cot\frac{\pi}{\omega}.
%\end{eqnarray}
By means of Eq. (\ref{eq:lambda}), it is then possible to derive the threshold for hyperbolic $\bm{x}_{p}$
%the threshold amplitude is obtained analytically as
\begin{eqnarray}
\label{eq:winding}
	A_{c}^{2}=\begin{cases}
		R^{-1} & \textrm{if $R>0$,}\\
		R^{-1}(1-\sin^{-2}\frac{\pi}{\omega}) & \textrm{otherwise.}
	\end{cases}
\end{eqnarray}
Equation (\ref{eq:winding}) allows one to label different unstable domains using the marginal stability condition $(\omega=2/m, A=0)$, with $m$ a positive integer. 
A numerical evaluation of  Eq. (\ref{eq:lambda}) (see Fig. \ref{eps:lambda}) identifies two distinctive regimes according to the ordering of $\omega$.
In the  high frequency regime with $\omega\gtrsim\mathcal{O}(10^{-1})$, unstable domains are distinguishable and the threshold amplitude $A_{c}$ varies rapidly with $\omega$. 
Meanwhile, the low frequency regime ($\omega\lesssim \mathcal{O}(10^{-1})$) is characterized by a dense spectrum corresponding to high order resonances with $m\gg 1$.
In this case, it is crucial to stress that the growth rate $\textrm{Im}\lambda$ exhibits insensitivity to $m$, and  $A_{c}$ piles up at a constant value $A_{c}=1$.

\begin{figure*}[!htp]
\centering
	\subfloat[\small{Adiabatic: $(q_{0}=3.2, p_{0}=0)$}]{
\label{eps:adiabatic}
\begin{minipage}[t]{0.32\textwidth}
\centering
\includegraphics[scale=0.40]{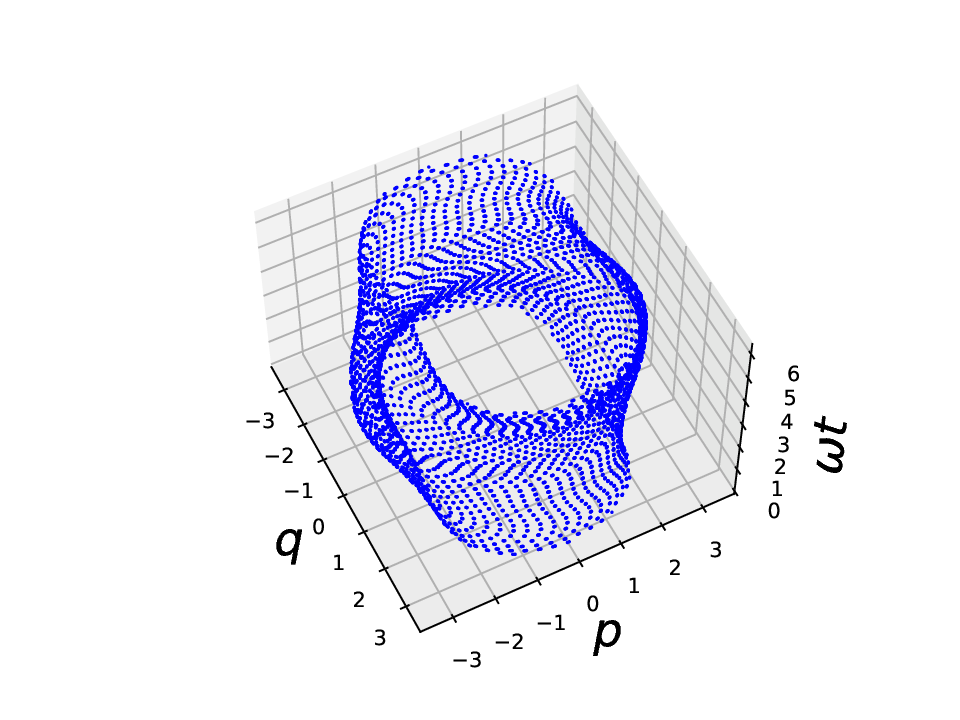}
\end{minipage}
}
	\subfloat[\small{Superadiabatic: $(q_{0}=2.5, p_{0}=0)$}]{
\label{eps:superadiabatic}
\begin{minipage}[t]{0.32\textwidth}
\centering
\includegraphics[scale=0.40]{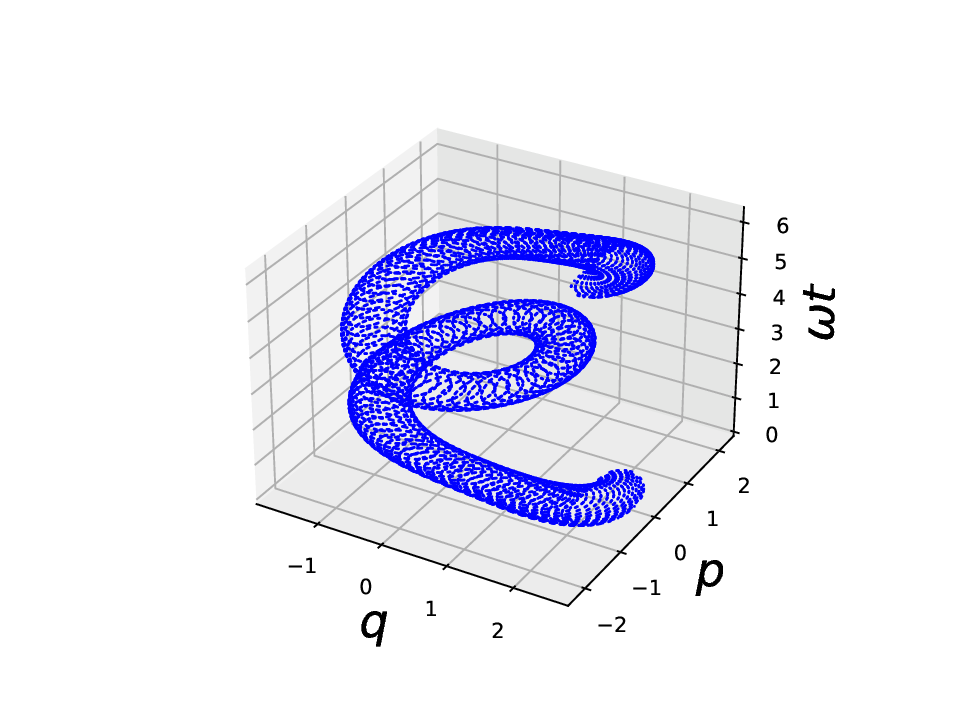}
\end{minipage}
}
	\subfloat[\small{Chaotic: $(q_{0}=1.0, p_{0}=0)$}]{
\label{eps:chaotic}
\begin{minipage}[t]{0.32\textwidth}
\centering
\includegraphics[scale=0.40]{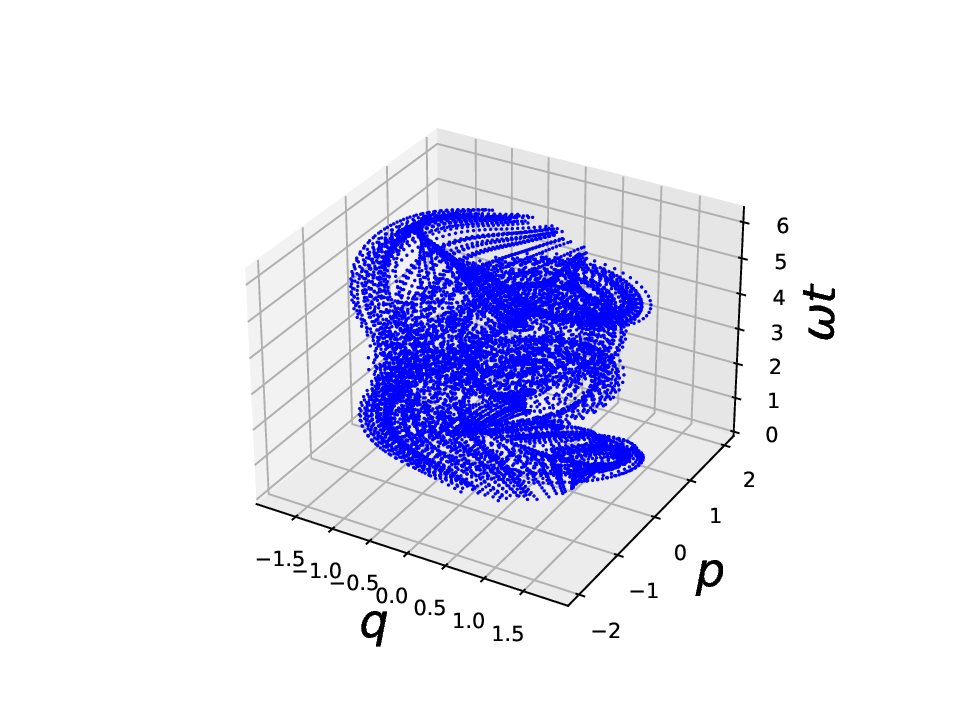}
\end{minipage}
}
	\caption{Comparison of the adiabatic, superadiabatic and chaotic orbits in the extended phase space. The orbits are plotted for $A=1.3$ and $\omega=0.5$ with different initial conditions $(q_{0}, p_{0})$.}
\label{eps:exorbits}
\end{figure*}

	The dynamical features of the system are closely connected to the type of fixed points, and, as illustrated in Fig. \ref{eps:orbits},  can be put into three categories. 
The adiabatic orbit  rotates with respect to the primary fixed point in  Poincar\'{e} map. 
	The associated Kolmogorov--Arnold--Moser (KAM) torus \cite{lichtenberg} is slightly deformed and remains intact topologically.  
It possesses the adiabatic invariance of $\bar{\mu}$, which can be derived using the perturbation theory and neglecting the resonance effect, as in Ref. \cite{taylor67}.
In regions of phase space where  resonances take place,  the original KAM torus will be destroyed and the perturbation series for $\bar{\mu}$ fails to converge \cite{dubin82}.
Nonetheless, a superadiabatic invariant \cite{rosenbluth72} may exist when a new KAM torus emerges, leading to the orbit rotating around a secondary  fixed point.
We refer to this type of orbit as the superadiabatic orbit. 
Chaotic orbit  is characterized by the destruction of separating KAM tori.
Quantitatively, the chaotic orbit in the present study is determined by the finite size Lyapunov exponent $\lambda_{L}$ \cite{aurell}.
Examples in Fig. \ref{eps:exorbits} show the topological differences between adiabatic, superadiabatic and chaotic orbits in the extended phase space.
The orbit in Fig. \ref{eps:chaotic}  displays chaotic behavior with $\lambda_{L}\simeq 0.014$.
%\end{linenumbers}

\begin{figure}[!htp]
\centering
	\subfloat[\small{$\omega=0.06$}]{
\label{eps:ome0d06_A1d01}
\begin{minipage}[t]{0.24\textwidth}
\centering
\includegraphics[scale=0.24]{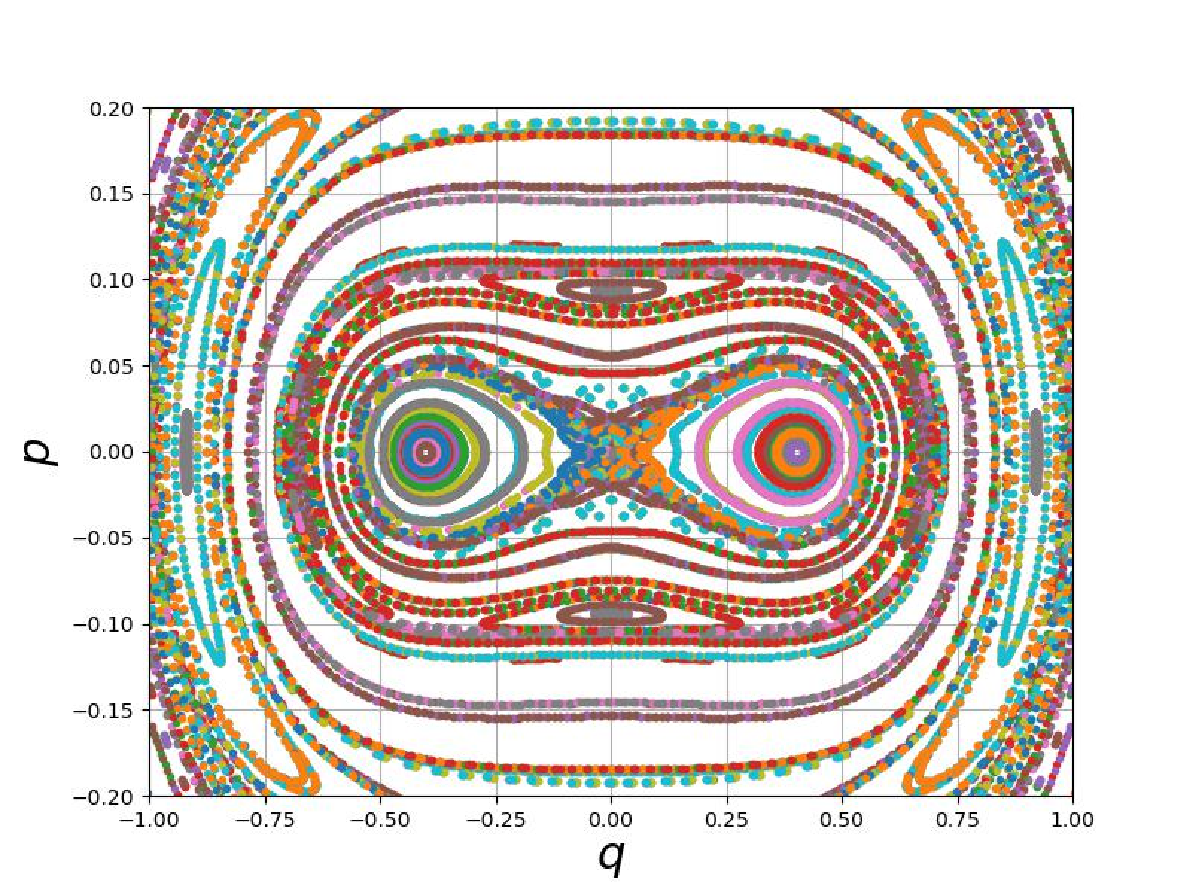}
\end{minipage}
}
	\subfloat[\small{$\omega=0.059$}]{
\label{eps:ome0d059_A1d01}
\begin{minipage}[t]{0.24\textwidth}
\centering
\includegraphics[scale=0.24]{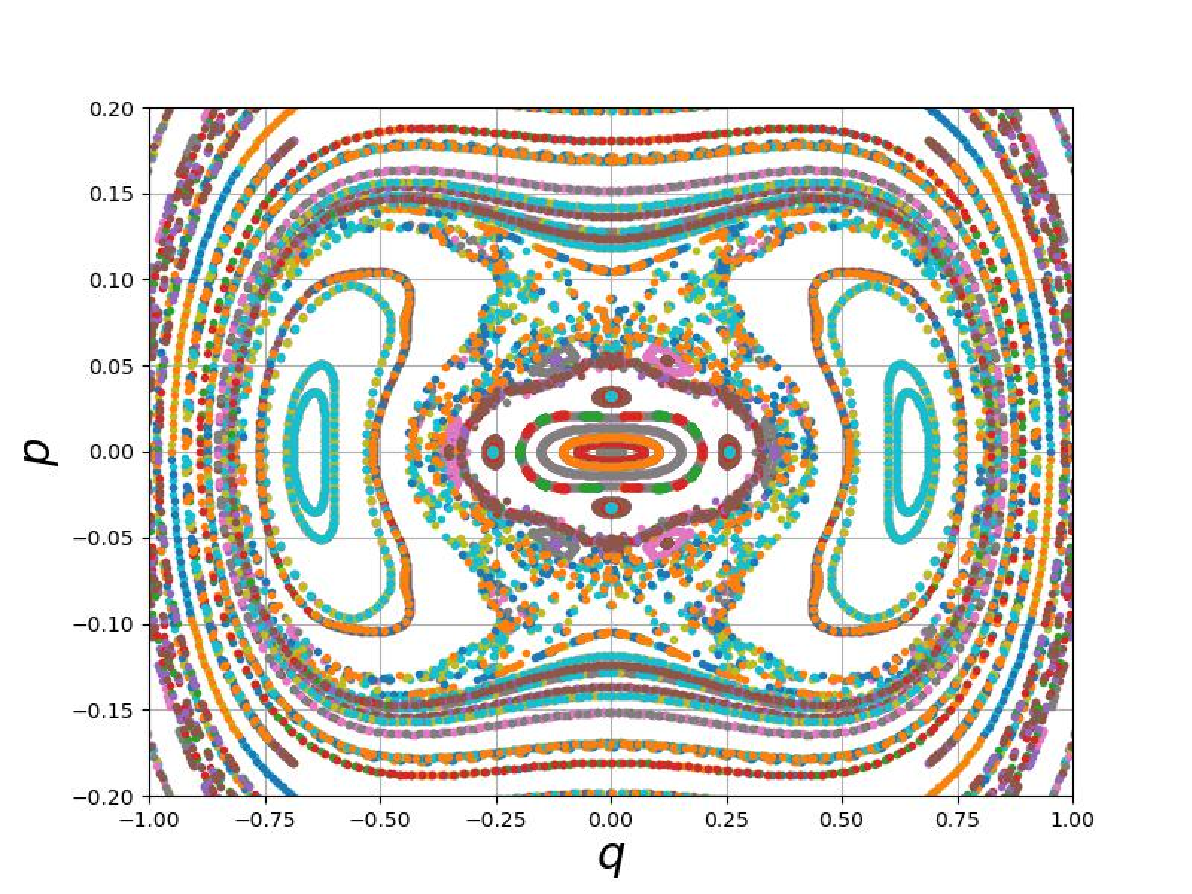}
\end{minipage}
}
	\caption{Phase space portrait of $\mathcal{M}_{T}$ for $A=1.01$. }
\label{eps:A1d01}
\end{figure}

%\begin{linenumbers}
%\section{Numerical Results}
%\label{sec:numerical}
Now we analyze the transition from adiabatic to superadiabatic/chaotic orbits by  scanning $A$ and $\omega$.
%The value of $\omega$ is scanned from $0.01$ to $1$. 
The white line in Fig. \ref{eps:lambda} demonstrates that there exists a quantitative, frequency-independent adiabaticity threshold $A_{c}\simeq 1$ for low frequency perturbations, below which the adiabatic invariant is preserved.
This effect thus ensures the general validity of gyrokinetic theory in the low frequency regime.
%This effect validates the applicability of gyrokinetic theory in the low frequency regime.
To understand the physical nature of this threshold, 
%To understand the underlying physics of this threshold, 
we notice that Eq. (\ref{eq:motion0}) can be rendered into a boundary layer problem  in the low frequency limit, yielding
\begin{eqnarray}
\label{eq:ilp}
	\omega^{2}\frac{d^{2}q}{d\theta^{2}}+q-A\cos \theta\sin q=0,
\end{eqnarray}
with $\theta=\omega t$. 
Equation (\ref{eq:ilp}) displays manifestly the vast disparity between the very short gyromotion time scale ($|\partial_{\theta}|\sim \omega^{-1}$), the wave-particle trapping time scale  ($|\partial_{\theta}|\sim \sqrt{A}/\omega$)  and the long time scale of the wave period ($|\partial_{\theta}|\sim 1$).
%Equation (\ref{eq:ilp}) is characterized by the vast disparity between the very short gyromotion time scale ($|\partial_{\theta}|\sim \omega^{-1}$), the wave-particle trapping time scale  ($|\partial_{\theta}|\sim \sqrt{A}/\omega$)  and the long time scale of the wave period ($|\partial_{\theta}|\sim 1$).
On the gyromotion time scale, $\cos \theta$ can be treated as a constant, and one can  easily show that the primary fixed point becomes unstable when $A>1$, in agreement with the previous stability analysis, Eq. (\ref{eq:lambda}).
In this case, the adiabaticity threshold $A_{c}=1$ could be interpreted as the condition that the wave-particle trapping frequency ($\simeq\sqrt{A}$) of resonant particles is comparable to the gyrofrequency ($=1$).
On the long time scale $\partial_{\theta}\sim 1$, we have $q-A\cos \theta\sin q\simeq 0$, then the secondary fixed points, which appear on time scales long compared to the wave period,  are only possible if $A\simeq |q/\sin q|\ge 1$.
%On the long time scale $\partial_{\theta}\sim 1$, we have $q-A\cos \theta\sin q\simeq 0$, thus the emergence of secondary fixed points, when viewed on time scales long compared to the wave period,  is possible only if $A\simeq |q/\sin q|>1$.
Additionally, numerical results suggest that $A_{c}=1$ also gives a quantitative estimate for the chaotic threshold in the low frequency regime, as seen in  Fig. \ref{eps:lambda}.
%This phenomenan may be physically interpreted as 
 Figure \ref{eps:ome0d06_A1d01} shows that a chaotic layer arises as the homoclinic tangle \cite{lichtenberg} formed in the proximity of hyperbolic primary fixed point.
But the onset of chaos  via the heteroclinic tangle \cite{lichtenberg}, originating from unstable hyperbolic secondary fixed points, is observed as well (see Fig. \ref{eps:ome0d059_A1d01}).
 %The strong sensitivity of the dynamics on $\omega$ is well evident by comparing Fig. (\ref{eps:ome0d06}) with (\ref{eps:ome0d059}).
%Based on the above results, one is led to conclude that the generality of gyrokinetic theory in the low frequency regime is not only because of the adiabatic invariance of $\bar{\mu}$,  but also due to the fact that the particle dynamics constitutes an initial layer problem.

%Based on the above results, one is led to conclude that the generality of gyrokinetic theory in the low frequency regime is not only because of the adiabatic invariance of $\bar{\mu}$,  but also due to the emergence of an initial layer problem in particle dynamics.
Based on the above results, one is led to conclude that the generality of gyrokinetic theory in the low frequency regime is not only because of the adiabatic invariance of $\bar{\mu}$,  but also a result  of the boundary layer problem arising in particle dynamics.
The latter guarantees the very existence of the quantitative, frequency independent threshold $A_{c}$, below which the adiabaticity is  generally preserved under low frequency perturbations.
%The latter guarantees that the adiabaticity is  generally preserved under low frequency perturbations, provided the wave amplitude is below the quantitative, frequency independent adiabaticity threshold $A_{c}=1$.
%The latter guarantees that the adiabaticity is  in general preserved under low frequency perturbations, provided the wave amplitude is below a frequency independent threshold  $A_{c}=1$.
%This may provide physical insight on the general validity of gyrokinetic framework in the low frequency regime.
%Therefore, the transition to chaos happens through the formation of either the homoclinic tangle around hyperbolic primarty fixed point, or the heteroclinic tangle arising from hyperbolic secondary fixed points.
%Above the threshold the system possessses both superadiabatic and chaotic behaviors.
%\end{linenumbers}

\begin{figure}[!htp]
\centering
	\subfloat[\small{$\omega=0.50$}]{
\label{eps:ome0d5_A1d4}
\begin{minipage}[t]{0.24\textwidth}
\centering
\includegraphics[scale=0.24]{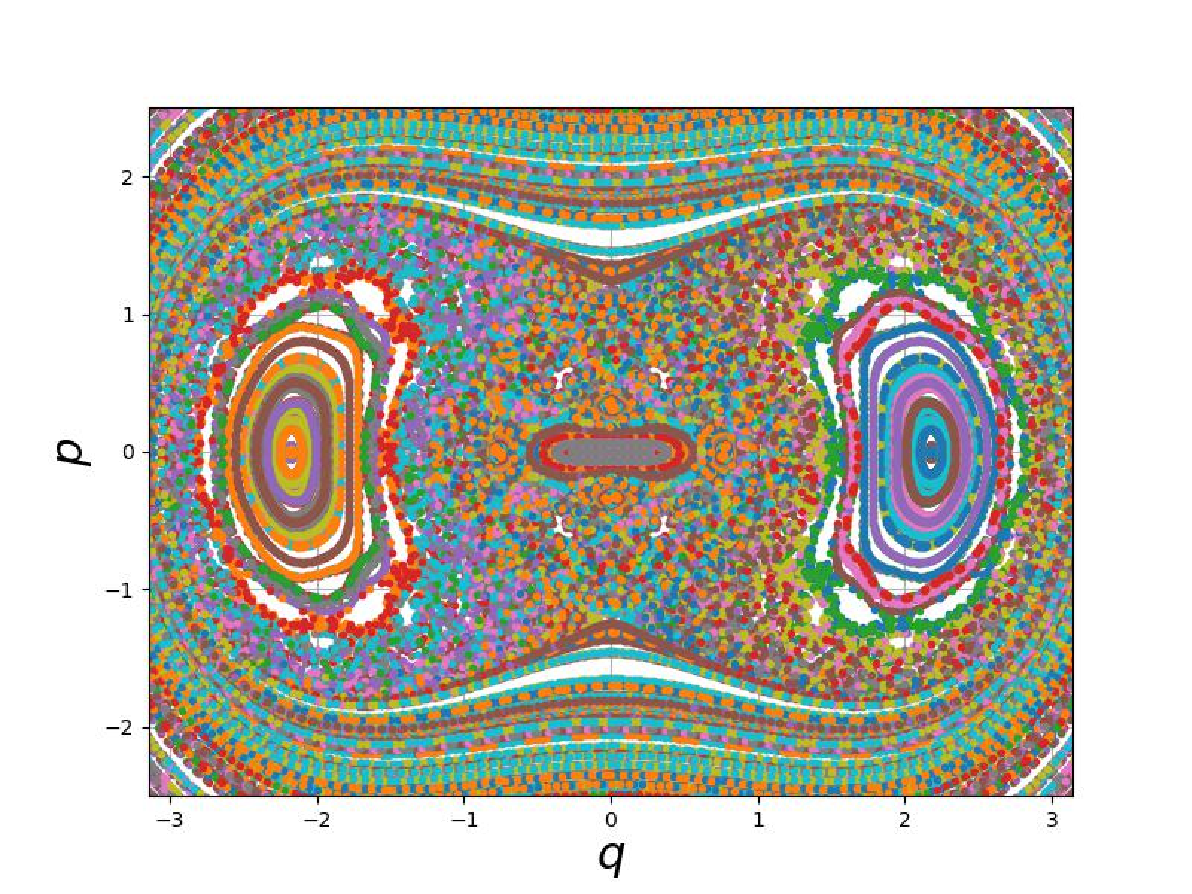}
\end{minipage}
}
	\subfloat[\small{$\omega=0.52$}]{
\label{eps:ome0d52_A1d4}
\begin{minipage}[t]{0.24\textwidth}
\centering
\includegraphics[scale=0.24]{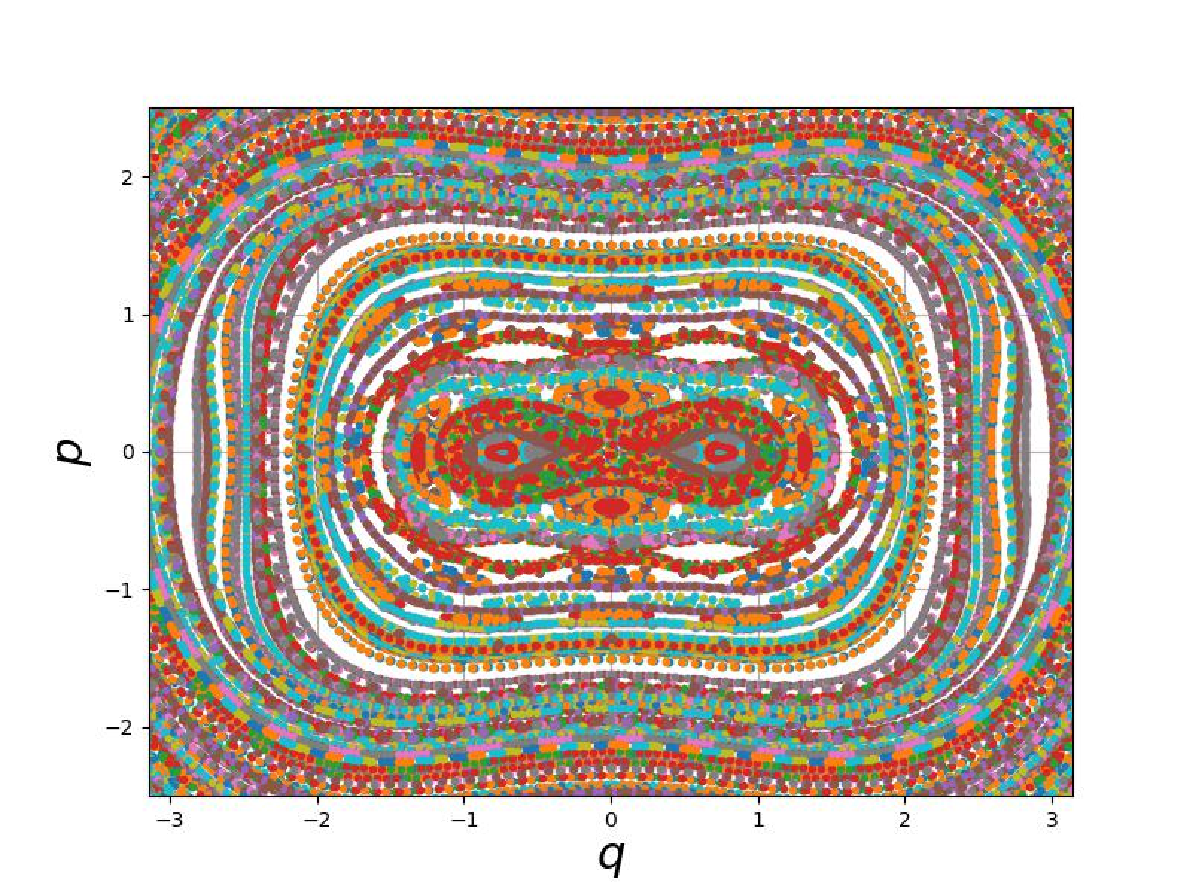}
\end{minipage}
}\\
	\subfloat[\small{$\omega=1.00$}]{
\label{eps:ome1d0_A1d4}
\begin{minipage}[t]{0.24\textwidth}
\centering
\includegraphics[scale=0.24]{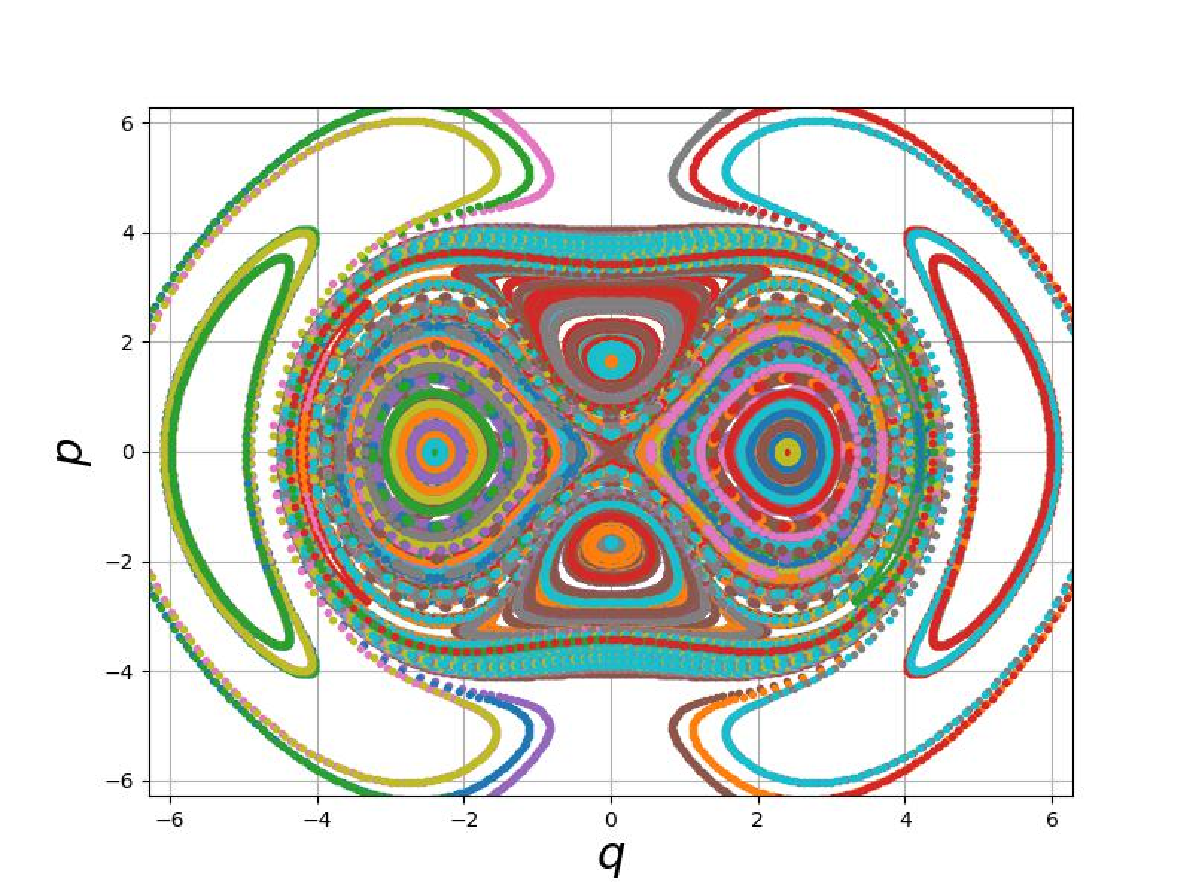}
\end{minipage}
}
	\subfloat[\small{$\omega=0.98$}]{
\label{eps:ome0d98_A1d4}
\begin{minipage}[t]{0.24\textwidth}
\centering
\includegraphics[scale=0.24]{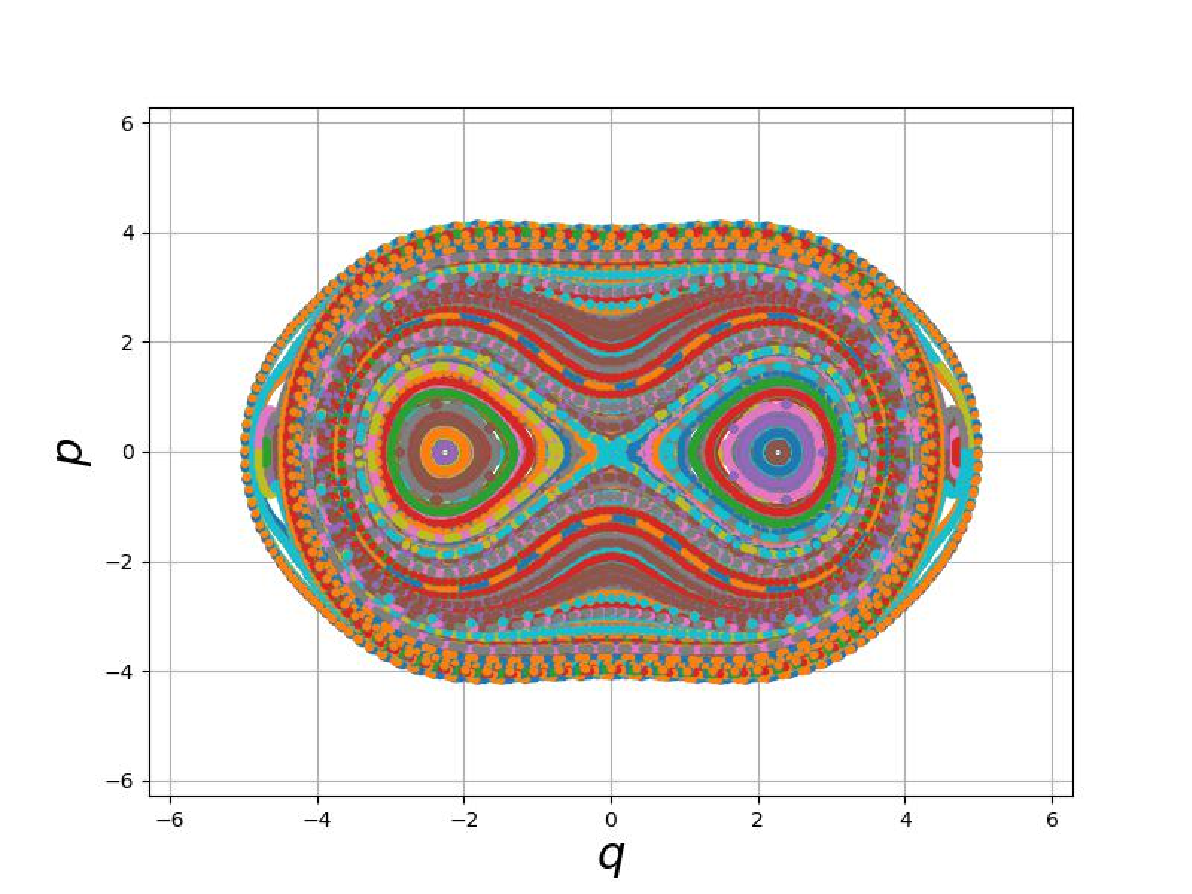}
\end{minipage}
}
	\caption{Phase space portrait of $\mathcal{M}_{T}$ for $A=1.4$. }
\label{eps:A1d4}
\end{figure}

%\begin{linenumbers}

The aforementioned separation of time scales does not apply for high frequency perturbations.
Accordingly,  Fig. \ref{eps:lambda} shows that both the superadiabatic and chaotic thresholds are highly sensitive to the wave frequency. 
%This is also consistent with 
Thus the adiabaticity is not universally preserved, as opposed to the low frequency case.
This difference then questions  the general validity of high frequency gyrokinetic theory assuming adiabatic orbits  \cite{chen83b,tsai84, qin99, qin00, kolesnikov,yu}.
%In a related vein,  the emergence of superadiabatic/chaotic orbits also vitiates another central hypothesis in high frequency gyrokinetic theory, which assumes that the existence of gyrocenter coordinates is independent of the wave frequency.
%In a related vein,  the breakdown of adiabaticity also vitiates the fundamental hypothesis that the existence of gyrocenter coordinates is independent of the wave frequency.
In a related vein,  the breakdown of adiabaticity vitiates another fundamental hypothesis that the existence of gyrocenter coordinates is independent of $\omega$  in the high frequency regime.
%However, this argument is also vitiated by the emergence of superadiabatic and chaotic orbits.
Specifically, noting that the perpendicular particle velocity in the present uniform plasma model satisfies  $v_{x}\propto p$ and $v_{y}\propto q$ \cite{taylor67}, the phase space portraits of $\mathcal{M}_{T}$  (shown in Fig.  \ref{eps:A1d4}) indicate that the near identity transformations between the particle and gyrocenter coordinates, which are expressed as asymptotic expansions in powers of the wave amplitude in Lie perturbation theory \cite{brizard07, krommes12, qin99, qin00}, will be invalidated by the presence of superadiabatic/chaotic orbits.
The high frequency gyrokinetic theory is therefore strictly valid only in the linear limit with $A\to 0^{+}$ \cite{chen83b, tsai84}.
%Therefore,  the high frequency gyrokinetic theory is strictly valid only in the linear limit with $A\to 0^{+}$.
 %When an arbitrary small but finite amplitude perturbation is considered, the linear high frequency gyrokinetic theory  will not adequately describe the particle dynamics near resonant points ($\omega=2/m$), due to the neglect of parametric resonance effect. 
 When an arbitrary small but finite amplitude perturbation is considered, the linear high frequency gyrokinetic theory  becomes inadequate for describing the particle dynamics near resonant points ($\omega=2/m$), due to the neglect of parametric resonance effect. 
In this context, the linear high frequency gyrokinetic theory could not be utilized to develop quasilinear or weak turbulence theories.

Unlike in the low frequency regime,  Fig. \ref{eps:lambda} demonstrates that even though high frequency perturbations may break the adiabaticity more easily, the onset of chaos deviates significantly from the superadiabatic threshold. 
In this case, one may attempt to develop a new reduced kinetic theory from the superadiabatic invariant.
However, the intrinsically nonperturbative nature of superadiabatic orbits and the associated complexity of phase space structures (cf. Fig. \ref{eps:A1d4})  make it difficult if not impossible to derive a comprehensive analytic description for the superadiabatic invariant. 
	This effect, thereby, precludes the possibility of new general reduced kinetic theories  built upon superadiabaticity, and  reinforces that the existence of invariant is not sufficient for reduced kinetic theories.

\begin{figure}[!htp]
\centering
	%\subfloat[Homoclinic tangle]{
	\subfloat[]{
\label{eps:SOSO}
\begin{minipage}[t]{0.24\textwidth}
\centering
\includegraphics[scale=0.30]{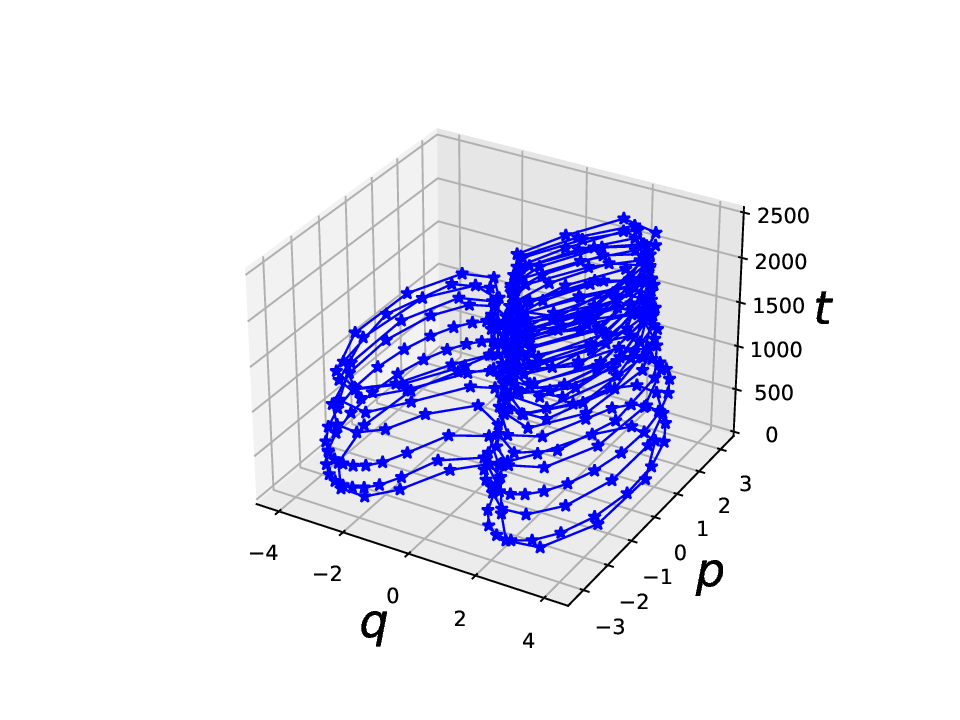}
\end{minipage}
}
	%\subfloat[\small{Heteroclinic tangle}]{
	\subfloat[]{
\label{eps:AOSO}
\begin{minipage}[t]{0.24\textwidth}
\centering
\includegraphics[scale=0.30]{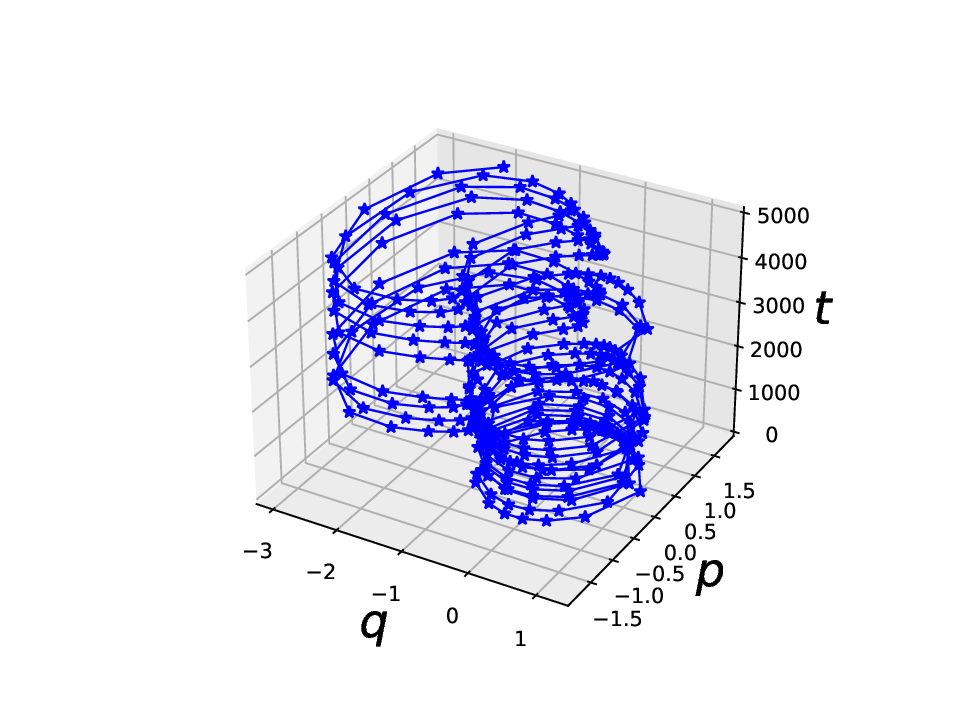}
\end{minipage}
}
	\caption{Typical evolution of marginally chaotic $\mathcal{M}_{T}$. The particle is randomly scattered into different types of orbits by homoclinic (a) or heteroclinic (b) tangles. }
\label{eps:jumporbits}
\end{figure}

Whilst it is well established that the Lie series \cite{dubin82}  of adiabatic/superadiabatic invariant will not converge for chaotic orbits  due to the presence of nonlinear resonance \cite{jaeger, chen01}, intuition may suggest that it might be possible to construct a gyrokinetic theory valid for a sufficiently long time via the gyro-averaging approach, if the orbit is only marginally chaotic.
Unfortunately, we find that once the chaotic threshold is exceeded,  homoclinic and heteroclinic tangles can scatter the charged-particle motion into different types of orbits without regularity, as shown in Fig. \ref{eps:jumporbits}.
In this scenario, the gyro-averaging  (which  refers to the integration along unperturbed adiabatic/superadiabatic orbits here) is ill defined.
It is thus impossible to properly account for the chaotic motion in both gyro-averaging and Lie perturbation approaches.
%\end{linenumbers}

%\begin{figure}[!htp]
%\vspace{-0.3cm}
%\setlength{\belowcaptionskip}{-0.1cm}
%\centering
%\includegraphics[scale=0.35]{stability2.eps}
	%\caption{Stability diagram of short wavelength CTEM, for $\tau=1$, $\epsilon=0.18$ and $s=0.8$. White is stable, \blue{and blue (red) represents unstable modes propagating in the ion (electron) diamagnetic direction.}}
%\label{eps:stability}
%\end{figure}

%\section{Discussion}
%\label{sec:may:01:16:04}

%\begin{linenumbers}

%\section{Conclusion}
%\label{sec:conclusion}
%In summary, through the simplest model, we have demonstrated that the gyrokinetic theory has general validity in low frequency regime  with $\omega\lesssim \mathcal{O}(10^{-1})$ if the nonlinearity parameter $A<1$, corresponding to the presence of well separated time scales for the cyclotron motion and wave-particle trapping.
In summary, by means of the simplest paradigm, the validity of gyrokinetic theory has been investigated.
%It is shown that, counter to frequent assertions, the formal validity of gyrokinetic theory rests not only on the existence of adiabatic invariant, but also on that the particle dynamics forms a  boundary layer problem.
Contrary to widely accepted assertions, it is shown that the existence of adiabatic invariant alone is not sufficient for the general validity of gyrokinetic theory. 
The particle dynamics must form a boundary layer problem to ensure the continued applicability of adiabaticity.
Within this framework, the gyrokinetic theory is valid for general low frequency fluctuations with amplitudes below the quantitative, frequency-independent threshold $A_{c}$.
In the high frequency regime, however, the adiabatic threshold displays sensitive dependence on the wave frequency and amplitude, which thereby calls into question the foundations of high frequency gyrokinetic theory. 
%Furthermore, because of the complex phase space behaviors, one cannot construct a new general reduced kinetic theory based on superadiabaticity.
Furthermore, because of the complex phase space behaviors, it is not feasible to construct a new general reduced kinetic theory based on superadiabaticity.
For these reasons, provided that the wave amplitude satisfies $A<A_{c}$, the gyrokinetic theory is reasonably applicable to the low frequency drift-Alfv\'{e}nic turbulence, whose wave frequency is typically $\mathcal{O}(10^{-2})$ smaller than the gyrofrequency \cite{chen16}.
%But the main possible concern remain possible concerns
%But possible concerns about the validity of gyrokinetic theory may raised in ion dynamics in the short wavelength electron temperature gradient turbulence \cite{chen21,tirkas},
Meanwhile, recalling the definition of $A$, it is worth mentioning the condition $A<A_{c}$ implies that the drift kinetic theory, which assumes  $k_{\perp}\to 0^{+}$ and $\omega\ll\Omega$, applies for arbitrary fluctuation amplitudes, consistent with previous results \cite{hazeltine,chen20}.
But still it is of great concern whether one can employ the gyrokinetic theory, for example,  in the vicinity of the separatrix of modern divertor tokamaks, where the radial wavenumber tends to diverge \cite{rozhansky}.
In this respect, all particles crossing the separatrix would be affected, including those away from the X-point.
%In this regard, all particles crossing the separatrix would be affected, including those away from the X-point.
%These findings provide useful insights into the physics basis of gyrokinetic theory.
%Finally, we remark that the present study is limited to the particle dynamics under perturbations propagating perpendicular to the magnetic field in a uniform plasma.
%It is obvious and desirable to extend the analyses to the realistic tokamak geometry. This will be the topic of a future investigation.
%\end{linenumbers}

%\section*{Acknowledgments}
%\label{sec:may:01:16:05}
H. C. wishes to thank Y. R. Lin-Liu and R. L. Miller for their proposal and help in the initial phase of this work.
This work was supported by the National Natural  Science Foundation of China under Grant Nos. U1967206, 12375213 and 12275071.
%This work was supported by the National Natural  Science Foundation of China under Grant Nos. U1967206 and 12275071.

%We acknowledge discussions with Z. Qiu, F. Zonca, E. Viezzer and M. Gracia-Munoz.
%This work was supported by National Natural Science Foundation of China under Grant No. 11905097, and the European Unions Horizon 2020 research and innovation programme (No. 805162).

\end{document}